# VST-SMASH: the VST Survey of Mass Assembly and Structural Hierarchy

Crescenzo Tortora[1]
Rossella Ragusa[1]
Massimiliano Gatto[1]
Marilena Spavone[1]
Leslie Hunt[2]
Vincenzo Ripepi[1]
Massimo Dall'Ora[1]
Abdurro'uf[3]
Francesca Annibali[4]
Maarten Baes[5]
Francesco Michel Concetto Belfiore[2]
Nicola Bellucco[6]
Micol Bolzonella[4]
Michele Cantiello[7]
Paola Dimauro[8]
Mathias Kluge[9]
Federico Lelli[2]
Nicola R. Napolitano[10]
Achille Nucita[11]
Mario Radovich[12]
Roberto Scaramella[8]
Eva Schinnerer[13]
Vincenzo Testa[8]
Aiswarya Unni[1]

[1] INAF–Capodimonte Astronomical Observatory, Naples, Italy
[2] INAF–Arcetri Astrophysical Observatory, Florence, Italy
[3] Department of Physics and Astronomy, The Johns Hopkins University, Baltimore, USA
[4] INAF–Bologna Astrophysics and Space Science Observatory, Italy
[5] Astronomical Observatory, Ghent University, Belgium
[6] Department of Physics and Astronomy, University of Padua, Italy
[7] INAF–Abruzzo Astronomical Observatory, Teramo, Italy
[8] INAF–Rome Astronomical Observatory, Italy
[9] Max Planck Institute for Extraterrestrial Physics, Garching, Germany
[10] Department of Physics, Federico II University, Naples, Italy
[11] Department of Mathematics and Physics, University of Salento, Italy
[12] INAF–Padua Astronomical Observatory, Italy
[13] Max Planck Institute for Astronomy, Heidelberg, Germany

The VLT Survey Telescope Survey of Mass Assembly and Structural Hierarchy (VST-SMASH) aims to detect tidal features and remnants around very nearby galaxies, a unique and essential diagnostic of the hierarchical nature of galaxy formation. Leveraging optimal sky conditions at ESO's Paranal Observatory, combined with the VST's multi-band optical filters, VST-SMASH aims to be the definitive survey of stellar streams and tidal remnants in the Local Volume, targeting a low surface-brightness limit of $\mu \sim 30$ mag arcsec$^{-2}$ in the $g$ and $r$ bands, and $\mu \sim 28$ mag arcsec$^{-2}$ in the $i$ band, in a volume-limited sample of local galaxies within 11 Mpc and the Euclid footprint.

## The low-surface-brightness realm

According to our current understanding of galaxy formation, cosmic structures form hierarchically. During the very early epochs, gas condenses, and the first stars are formed within the primordial dark matter perturbations (Blumenthal et al., 1984; White & Frenk, 1991). These first galaxies then merge, resulting in larger and larger galaxies, which in turn undergo physical transformations governed by the wide range of different environments they live in. Major mergers (for example involving galaxies of equal mass) are the most spectacular manifestation of hierarchical assembly, but only a small fraction of galaxies are involved in such catastrophic events that are expected to destroy pre-existing discs. Instead, minor mergers (for example involving galaxies having a 1:10 ratio of mass) are expected to be more common (see, for example, Cole et al., 2000). If satellite galaxies, orbiting the stellar body of the massive central companion, are tidally disrupted, a variety of tidal structures, such as shells, streams and plumes, which are a direct 'smoking gun' of hierarchical assembly, should emerge (see, for example, Cooper et al., 2010). Within the standard ΛCDM cosmological model (Springel et al., 2008), such features can provide important constraints on the formation of the stellar halo (Bell et al., 2008), gravitational potential (Bovy et al., 2016) and dark matter substructures (Sandford et al., 2017). To better understand the hierarchical structural assembly, it is necessary to systematically study the remnants of such interactions. This can be achieved via deep imaging of nearby galaxies and their surroundings to address the following still-open questions: What is the frequency of tidal interactions and galaxy accretion via minor mergers in the very Local Universe? What are the properties of their stellar populations? Are observations consistent with numerical simulations and the current paradigm for galaxy formation?

## The state of the art

Tidal structures are extremely faint, with very low surface brightness (LSB; $\mu_V \geq 27$ mag arcsec$^{-2}$; Johnston et al., 2001), and therefore have been historically difficult to observe. With Gaia and the Dark Energy Survey, the study of stellar streams has entered a golden age (see, for example, Belokurov et al., 2017; Shipp et al., 2018). Nevertheless, the challenge now is to go deeper and beyond the Local Group (LG). The Milky Way and its companions by themselves cannot provide a representative picture of the hierarchical structure assembly of the Universe. Outside the LG, the first systematic ground-based searches have so far been limited to only a few galaxies, relying on wide-band photometry and therefore lacking the colour information needed for a characterisation of the stellar populations (see, for example, Martinez-Delgado et al., 2010). Space-based HST observations (for example, Radburn-Smith et al., 2011) incorporate different broad-band filters, providing excellent spatial resolution but with a very limited field of view (FoV). Neither type of approach can provide a complete picture of the overall statistics of such features. The limitation of the FoV has been overcome in the last 10 years, mainly in the northern cap, by the surveys MATLAS (Bílek et al., 2020), ELVES (Carlsten et al., 2022), SSH (Annibali et al., 2020) and LIGHTS (Trujillo et al., 2021).

## The VST: a seeker of faint structures

Next-generation facilities, such as Vera C. Rubin Observatory or ESA's Euclid mission, thanks to their large FoV and superb spatial resolution, will improve our understanding of galaxy assembly in the local Universe. Unfortunately, Vera C. Rubin Observatory's Legacy Survey of Space and Time (LSST) will reach the required surface brightness (SB) limits only after about 10 years of observations and Euclid will observe in three near-infrared (NIR) bands and mounts only a single broad-band optical filter. A mission planned to achieve these scientific goals, reaching an unprecedentedly ultra-low SB, known as ARRAKIHS[1], will only be launched in the early 2030s. Therefore, the VLT Survey Telescope (VST) is currently the ideal instrument with which to study tidal features and hierarchical structuring. Thanks to the unique combination of exquisite seeing and dark skies at the Paranal site and OmegaCam's FoV of 1 square degree, the VST can reveal the very faint structures around galaxies in the Local Volume. It has already proven its ability to study such structures within the VST Early-type GAlaxy Survey (VEGAS; Capaccioli et al., 2015; Iodice et al., 2021) and the Fornax Deep Survey (FDS; Iodice et al., 2019; Spavone et al., 2020), which obtained maps of the SB of early- and late-type galaxies, detecting the signatures of diffuse stellar components with azimuthally averaged SB depths of $\sim 29$–$30$ mag arcsec$^{-2}$ in the $g$ band (see also Iodice et al., 2019; Ragusa et al., 2021, 2022).

### VST-SMASH: a game changer in the Local Volume

Until now a systematic, homogeneous census of features reaching $\mu_g \sim 29$–$30$ mag arcsec$^{-2}$ around very nearby galaxies in the southern sky was missing. The VST Survey of Mass Assembly and Structural Hierarchy (VST-SMASH) will fill this gap, by observing, in the $g$, $r$ and $i$ bands, a sample of 27 local galaxies within the Euclid footprint in a distance-limited (< 11 Mpc) sample (Karachentsev, Makarov & Kaisina, 2013). These data will provide an unprecedented look at the hierarchical galaxy assembly, complementing the VEGAS and FDS datasets with a wide range of galaxy types (from dwarfs to massive spirals). Comparing the specifics of VST-SMASH with past surveys, we find that with the exception of SSH, which explores the same volume but focuses on dwarf galaxies, and ELVES, the previous surveys have observed galaxy samples at greater distances than ours. VST-SMASH is observing a sample as large as that observed by ELVES, including less massive systems.

### The survey objectives

To obtain the most complete characterisation of the mass assembly at $D < 11$ Mpc, we select a volume-limited galaxy sample, which includes objects of different types, masses and sizes, exploiting the VST and OmegaCAM characteristics. We reach an SB of about 30, 30 and 28 mag arcsec$^{-2}$, in the $g$, $r$ and $i$ bands respectively, with an exposure time of 2.5 hours in $g$ and $r$ and of 2 hours in $i$, and azimuthally averaged profiles. Observing both the fields around the target and a contiguous one for a proper background subtraction (Spavone et al., 2018), we can probe their halos, depending on their distances, up to about 250 kpc.

### Sample selection and data reduction

From the 869 galaxies in the distance-limited (< 11 Mpc) Updated Nearby Galaxy Catalogue (Karachentsev, Makarov & Kaisina, 2013), we have selected all the galaxies at declinations ≤ 5 deg, with a Holmberg diameter $a_{26} \geq 5$ arcmin and within the Euclid wide-survey footprint. We have purposely excluded: i) the largest galaxies (LMC, SMC, Sag DSph) and the Milky Way companions because they have already been extensively studied; ii) NGC 3115 since it is already observed by the VST; and iii) galaxies with overlapping and overwhelmingly bright stars. We end up with 27 galaxies. The sample spans a wide range of mass ($10^9$ – $2 \times 10^{11}$ $M_\odot$), HI gas mass ($10^8$ – $10^{10}$ $M_\odot$), types, orientations and environments. The target galaxies are also characterised by a wide range of apparent sizes: $a_{26}$ in the range of 5–40 arcminutes. The VST, with its wide FoV, is the perfect facility to efficiently observe such extended galaxies and their surroundings. Data are reduced with the AstroWISE pipeline (McFarland et al., 2013), which performs instrumental corrections, background subtraction and calibrations, and creates the final mosaics. Table 1 presents the galaxy sample.

Table 1. Data sample. Data on Coordinates, distance, Holmberg diameter and mass are extracted from Karachentsev, Makarov & Kaisina (2013). In the last column, we show the sky area observed around galaxies with complete VST-SMASH data in all the bands.

| Galaxy | R. A. (deg) | Dec. (deg) | Distance (Mpc) | $a_{26}$ (arcmin) | log($M/M_\odot$) | Area (sq. deg.) |
|---|---|---|---|---|---|---|
| NGC 0024 | 2.485 | –24.963333 | 9.9 | 7.24 | 10.32 | 0 |
| NGC 0045 | 3.51625 | –23.182222 | 9.2 | 8.51 | 10.45 | 0 |
| NGC 0055 | 3.785417 | –39.220278 | 2.13 | 37.15 | 10.15 | 0 |
| NGC 0247 | 11.784583 | –20.76 | 3.65 | 25.12 | 10.44 | 0 |
| NGC 0253 | 11.892917 | –25.292222 | 3.94 | 37.15 | 11.24 | 0 |
| NGC 0300 | 13.722917 | –37.6825 | 2.15 | 25.7 | 10.18 | 0 |
| NGC 0625 | 23.770833 | –41.436389 | 3.89 | 7.08 | 9.01 | 0 |
| ESO 115-021 | 39.4375 | –61.341111 | 4.99 | 7.24 | 9.5 | 0 |
| ESO 154-023 | 44.21 | –54.573056 | 5.55 | 8.32 | 9.62 | 0 |
| ESO 300-014 | 47.4075 | –41.030556 | 9.8 | 7.08 | 10.0 | 0 |
| NGC 1291 | 49.3275 | –41.108056 | 8.8 | 14.45 | 9.77 | 0 |
| NGC 1313 | 49.564167 | –66.4975 | 4.07 | 12.59 | 10.44 | 0 |
| NGC 1744 | 74.9925 | –26.026667 | 10.0 | 6.92 | 10.38 | 2 |
| NGC 3109 | 150.78 | –26.16 | 1.32 | 19.95 | 9.37 | 2 |
| Sextans A | 152.753333 | –4.692778 | 1.32 | 5.89 | 8.4 | 1 |
| NGC 3621 | 169.567083 | –32.811667 | 6.7 | 12.3 | 10.73 | 2 |
| NGC 5068 | 199.730417 | –21.039167 | 5.45 | 10.0 | 9.94 | 2 |
| NGC 5236 | 204.250417 | –29.867778 | 4.92 | 18.62 | 11.32 | 2 |
| NGC 5253 | 204.9825 | –31.64 | 3.56 | 6.92 | 8.91 | 1 |
| IC 5052 | 313.025833 | –69.203889 | 6.03 | 8.13 | 9.97 | 0 |
| NGC 7090 | 324.119167 | –54.557222 | 6.7 | 10.23 | 10.26 | 0 |
| IC 5152 | 330.674583 | –51.295278 | 1.97 | 7.08 | 8.94 | 0 |
| NGC 7462 | 345.696667 | –40.835 | 10.1 | 5.5 | 10.06 | 0 |
| IC 5332 | 353.614583 | –36.101667 | 7.8 | 8.32 | 10.24 | 2 |
| NGC 7713 | 354.0625 | –37.938889 | 7.8 | 6.76 | 10.11 | 1 |
| UGCA 442 | 355.941667 | –31.959167 | 4.27 | 6.31 | 9.16 | 0 |
| NGC 7793 | 359.455833 | –32.59 | 3.91 | 14.13 | 10.28 | 0 |

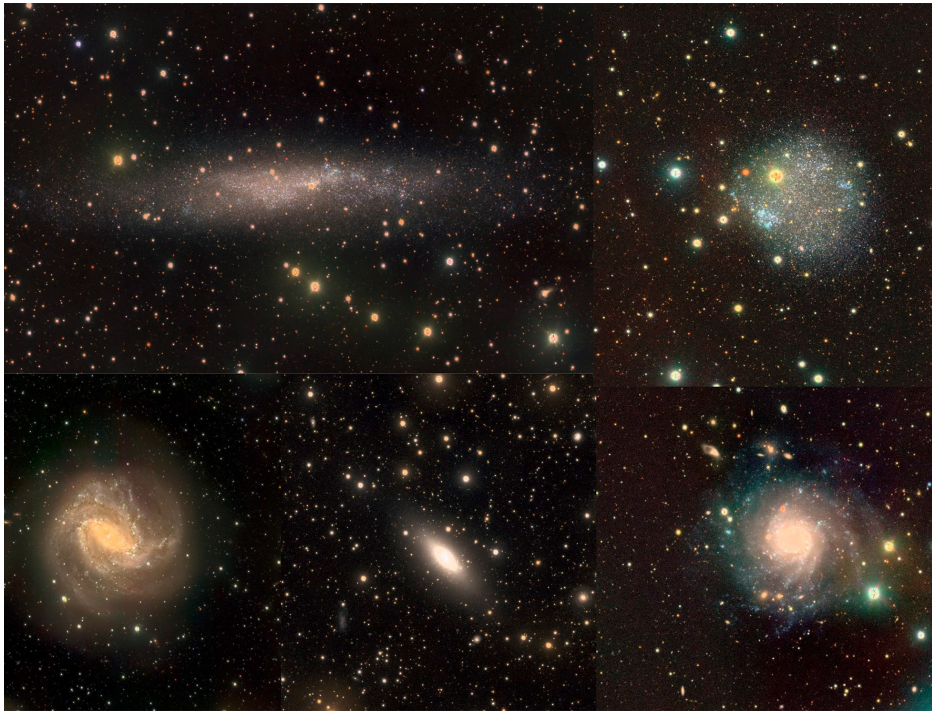

Figure 1. VST *gri* composite images of NGC 3109, Sextans A (top), NGC 5236, NGC 5253, IC 5332 (bottom).

Scientific objectives

The main aims of the survey are the following:
1. Detection and statistics of tidal features, LSB galaxies and star clusters in the outskirts of galaxies, up to galactocentric distances of about 50–250 kpc, based on their distance. 1D and 2D models of the light distribution are derived by fitting the isophotes and the latter are then subtracted from the parent images. This will allow us to detect any asymmetry in the galaxy's outskirts and the remnants of past accretion/ merging events. Using multiple analytical profiles, we can then set the scales of the different components in the light distribution, identifying the regions where the diffuse stellar envelope plus LSB features dominate over the in-situ populations. For the two closest galaxies (NGC 3109 and Sextans A), the analysis is performed via resolved star counts (for example, Annibali et al., 2020).
2. Determination of colour maps and a probe of the stellar halo populations by tracing galaxy SB to the outskirts. Colour profiles and possible break radii provide unique information about the interplay between internal galaxy processes and the assembly in the outskirts. In-situ processes vs ex-situ accretion leave a signature in the colour gradients, interpreted via hydrodynamical simulations (for example, Cook et al., 2016).
3. Characterisation of the stellar populations using *g*, *r*, *i* photometry of the main galaxy body and of the galaxy peripheries, to disentangle accreted stars from the in-situ populations. More precise stellar parameters, primarily stellar mass, will be constrained by fitting the spectral energy distribution (Abdurro'uf et al., 2021). Euclid will provide deep NIR bands. Meanwhile, we will complement our optical data with literature photometry in the ultraviolet and NIR, using the extended wavelength range to constrain stellar populations in the main galaxy bodies.
4. Inventories of star clusters and dwarf companions following the methodology described by Cantiello et al. (2020) and Venhola et al. (2018), respectively, which relies on automatic detection tools that take into account background/foreground contamination and are used to map their number density, structural properties and stellar population properties. These small stellar systems are the main contributors of extended stellar halos and the 2D density map of star clusters can provide hints about the galaxy mass assembly. For example, the adopted observing strategy will enable the detection and analysis of the old halo globular cluster (GC) populations in all targets, extending well beyond the peak of the GC luminosity function. Detection of satellites in both dwarfs and massive spirals would observationally test the self-similarity of the hierarchical formation process at all scales.
5. Comparison with mock galaxies. Large-volume hydrodynamical simulations (for example, TNG50 or NewHorizon; Pillepich et al., 2019; Dubois et al., 2021), including not only gravity but also stellar and gas physics and feedback from supernovae and AGN, provide the most efficient way to quantitatively interpret the statistics of tidal features and companions, and their properties (see, for example, Martìnez-Delgado et al., 2023), exploring the galaxy-halo connection and putting under a magnifying glass our understanding of the cosmological model and the galaxy mass assembly. In turn, such deep data can constrain the 'subgrid' models implemented in the simulations. Synthetic multi-band images of mock galaxies will be generated using the SKIRT radiative transfer code, which takes into account stellar emission by various stellar populations, absorption and scattering by interstellar dust (Baes et al., 2024). Such mock observations with realistic and physically motivated structures, stellar populations and statistics of LSB features will provide a sensitive testbed for galaxy evolution studies.
6. Complementary archival WISE, GALEX, $H_2$, ALMA, SPITZER/IRAC, HI, JWST data, which probe wide spectral and resolution ranges, can be used to determine star formation rates, stellar masses, rotational velocities, and HI and $H_2$ gas masses. Comparing our deep optical observations with such data (in particular the high-resolution ones from ALMA, JWST and HST), will allow us to constrain the interplay between the small-scale physics of gas and star formation and galaxy-scale properties, like galaxy morphology and stellar populations in the local volume.

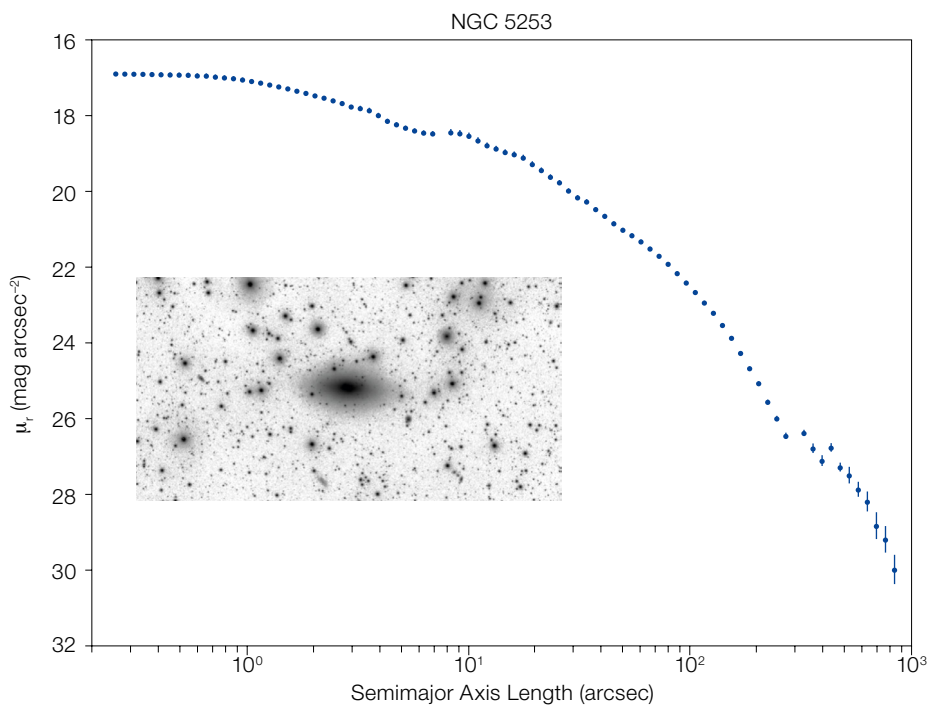

Figure 2. *r*-band azimuthally averaged surface brightness profile of NGC 5253. Surface brightness is plotted against the semimajor axis length in arcseconds. Our *r*-band photometry reaches a depth of 30 mag arcsec$^{-2}$. Inset: A cutout of the r-band image of the galaxy, aligned along the semimajor axis, is displayed. The size of the cutout matches the major-axis length corresponding to the outermost point in the SB profile.

7. As a byproduct, the large FoV will also enable the studies of stellar and structural parameters of the background galaxies in terms of mass, redshift and environment, leading to the discovery of gravitational lenses, quasars, and clusters.

### Status of the survey and first observations

At the time of writing, observations for the VST-SMASH programmes have run for two semesters, and observations for a third semester have been granted. Nine out of 27 targets have complete observations in the *g*, *r* and *i* bands, and are listed in Table 1, with their coordinates, distances, Holmberg radii, total masses and the sky area observed around that specific target. Figure 1 shows a montage of the colour composite images for: the two irregular galaxies NGC 3109 and Sextans A, the only two galaxies resolved in stars among our targets, respectively on the top left and top right panels; the face-on barred spiral galaxy NGC 5236 (the well-known Southern Pinwheel galaxy) with the companion starburst galaxy NGC 5253, on the bottom left and bottom central sides; and the intermediate spiral galaxy IC 5332 on the bottom right panel. The first data analysis has confirmed that our observing strategy, under the required seeing and sky conditions, has reached the requested depths. In particular, our SB azimuthally averaged profiles reach depths of around 29.5–30 mag arcsec$^{-2}$ in the *g* and *r* bands. In Figure 2 we show, as an example, the *r*-band surface brightness profile of NGC 5253. During the current semester, we are planning to finish the incomplete galaxies and add more observations for some other targets, aiming at completing the programme in the next few semesters.

### Conclusions

Thanks to the VST's capabilities, and the good sky conditions at ESO's Paranal Observatory, VST-SMASH is collecting exceptional data in the *g*, *r* and *i* bands, providing the most homogeneous legacy survey of stellar streams and tidal remnants in the very Local Volume for years to come, observing a volume-limited sample of galaxies at $D \leq 11$ Mpc. VST-SMASH is obtaining an image depth that LSST will reach only close to the end of its operations, owing to its observing strategy, and will contribute to the science that will be performed with the NIR Euclid imaging, providing the deep optical counterpart for this set of galaxies. Our dataset and the analysis we are performing will represent a fundamental benchmark for simulations and our understanding of hierarchical mass assembly.

### Links

[1] ARRAKIHS: https://www.cosmos.esa.int/web/call-for-missions-2021/selection-of-f2